\documentclass[10pt,a4paper,twoside,mathserif]{article}
\usepackage{epsfig} 
\usepackage{graphicx}
\usepackage{baltlat6}
\usepackage{array}
\usepackage{here}
\usepackage[koi8-r]{inputenc}
\usepackage{natbib}
\usepackage{url}
\usepackage{hyperref}
\newcommand{\kms}{\mbox{km~s}^{-1}}

\newcommand{\kpc}{\ensuremath{\rm kpc}}
\renewcommand{\mag}[1]{^{\rm m}\!\!\!#1\,}
\newcommand{\magdot}[1]{^{\rm m}\!\!\!#1\,}

\newcommand{\Lsun}{\ensuremath{\rm L_\odot}}

   \newcommand{\aj}{AJ}                      
   
   \newcommand{\apj}{ApJ}                      
   \newcommand{\apjl}{ApJL}                    

   \newcommand{\aap}{A\&A}                      
   
   \newcommand{\mnras}{MNRAS}
   \newcommand{\na}{New Astronomy}
   \newcommand{\nar}{New Astronomy Review}

   \newcommand{\pasp}{Publications of the ASP}
   
   \newcommand{\rmxaa}{Rev. Mex.}

\begin{document}
\ \
\vspace{0.5mm}
\setcounter{page}{0}

\titlehead{Baltic Astronomy, vol.\, ,     , 2014}
\titleb{THE HALF-CENTURY HISTORY OF STUDIES OF ROMANO'S STAR}

    \begin{authorl}
    \authorb{Olga Maryeva}{~} 
    \end{authorl}
    
    \begin{addressl}
   \addressb{~}{Special Astrophysical Observatory of the Russian Academy of Sciences, Nizhnii Arkhyz, 369167, Russia;
    olga.maryeva@gmail.com}
    \end{addressl}
    
\submitb{               } 

\begin{summary} 
We present a
short review of the observations and simulations of one of the most
interesting massive stars, V532, or Romano's star, located in the M33 galaxy. 
\end{summary}


\resthead{History of Romano's star studies}
{Olga Maryeva}
   
\sectionb{1}{INTRODUCTION}

            Luminous blue variables (LBVs) are rare objects of very
            high luminosity ($\sim10^6 L_{\odot}$) and mass loss
            rates  ($10^{-5}{M}\div10^{-4} {M}_\odot\mbox{yr}^{-1}$), low wind velocities,  exhibiting strong
            irregular photometric and spectral variability ~\citep{Conti,HumphreysDavidson,Humphreys2014}.
            They are generally believed 
            to be a relatively short evolutionary stage in the 
            life of a massive star, marking the transition from 
            the Main Sequence toward Wolf-Rayet (WR) stars. 
            However, recent studies indicate that progenitors of several 
            supernovae underwent LBV-like eruptions.            
            These studies support the view that at least some LBV
            stars are the end point of the evolution  but not a transition phase. 
            LBVs are rare objects,  observations of whose in the Galaxy are inevitably connected with 
            difficulties in determination of the distance and interstellar extinction. Hence, studying these rare objects in nearby galaxies is 
            potentially more prospective. 
            Therefore, investigation of the extragalactic star V532\footnote{Romano's star  (V532 or GR290) with
           $\alpha$=01:35:09.712, $\delta$=+30:41:56.55 according   to SIMBAD data base.}, which is now classified as
          LBV/post-LBV star and shows late-WN spectrum, is very important for our understanding
                of evolution of massive stars in general.

\sectionb{2}{PHOTOMETRIC OBSERVATIONS}

     Photometric observations of V532 had been started by Giuliano Romano in
     Asiago observatory (Italy) in  the early 1960s.
     \citet{romano} demonstrated a light curve from which one may see that stellar magnitude of V532 
     star irregularly varied between 16$\mag{.}7$ and 18$\mag{.}1$.
     G.~Romano classified V532 as a variable of the Hubble-Sandage type by
     the shape of the light curve and its colour index \citep{romano}.

     Photometric investigations were continued by \citet{Kurtev2001}.
     \citet{Kurtev2001} combined data by G.~Romano with their series of
     photometric observations, carried out during 8 years. 
     They found that during the whole period of
     study Romano's star displayed two maxima of brightness. The first  was
     near 1970 year and the second -- in early 1990s. Moreover
     \citet{Kurtev2001} discovered short-timescale variability with
     amplitude $\sim$0$\mag{.}$5 which  is typical for an LBV star.

     \citet{Zharova2011} investigated photometric variability of V532 using
     the Moscow collection of photographic plates. 
     They combined these data with new data obtained at  Russian 60-cm Zeiss
     telescope of Sternberg Astronomical Institute and  Russian   1-m Zeiss telescope
     of Special Astrophysical Observatory (SAO) and with photometric data published
     earlier by \citet{romano,Humphreys80,viotti2006}. %
     Thus \citet{Zharova2011} constructed   the most comprehensive light curve which covers 50 years
     of observations. The light curve shows that V532 exhibits irregular
     light variations with different amplitudes and time scales. Generally,
     the star shows large and complex wave-like variations, with duration
     of the waves amounting to several years.  
Four maxima of the waves were
     observed \citep{Zharova2011}. 
     In general, its variability is irregular,
     with the power spectrum fairly approximated by a red power-law spectrum
     \citep{PashaLBV}. 

     In late 2010 small brightening of Romano's star was detected, the star
     again reached 17$\mag{.}$8 in $V$-band \citep{Viotti2014ATel}.
     \citet{Viotti2014ATel} reported that after a moderate luminosity maximum at the beginning of
     2011, the Romano's star  has reached a new deep minimum with V=18$\mag{.}$7
     and R=18$\mag{.}$4 in December 2013, which appears to be the deepest so far
     recorded in its known light curve  history.

     Moreover photometric observations of V532 were carried out in the infrared range. 
     Stellar magnitudes are J=16$\mag{.}$8, H=16$\mag{.}$87 and  K=16$\mag{.}$8 according to 2MASS survey \citep{2MASS},
     16$\mag{.}$3 and  15$\mag{.}$9 in 3.6 $\mu$m and 4.5 $\mu$m bands, respectively, according to the data of { Spitzer space telescope} \citep{Spitzer1}.



\sectionb{3}{SPECTRAL OBSERVATIONS}

     Spectral observation of Romano's star began only in 1992. T.~Szeifert obtained the 
     first optical spectrum of Romano's star at the 3.5-m Calar Alto telescope \citep{szeifert}. 
     Red range of the spectrum is shown in \citet{szeifert} (Figure~5 in the original article). 
     These spectrum was described by T.~Szeifert as ``Few metal lines are visible, although 
     a late B spectral type is most likely''. 
     

    Next spectrum was obtained by \citet{Olga97} on the Russian 6-m
    telescope. \citet{Olga97} classified    V532 as a WN star candidate based on the  similarity of its
    spectrum  to the one of MCA~1B. MCA~1B is also located in M33 galaxy
    and  was classified as Ofp/WN9 \citep{MCA1B,MCA1BSmith}.     
    Since 1998 year 
    regular observations of V532 are being carried out in SAO \citep{fabrika,Sholukhova2011}.   
   \citet{Sholukhova2011} published detail series of spectral observations.


       Since 2003 spectral observations are being carried by Italian astronomers on telescopes of Cima Ekar (Asiago) and Loiano (Bologna) observatories \citep{polcaro,viotti2006,polcaro10}. Moreover spectra of V532 at two important extrema -- at the  minimum of brightness in 2008 and at 
       the moderate luminosity maximum of 2010 -- were described by \citet{polcaro10}, \citet{me} and \citet{ClarkLBV2012}, respectively.
       
       

 

\sectionb{4}{CIRCUMSTELLAR NEBULA}

Nebular lines of [OIII]~$\lambda\lambda 4959, 5007$,[NII]~$\lambda\lambda 6548-83$, [ArIII]~$\lambda7135 $   
and [FeIII]~$\lambda\lambda 4658,4701$ are clearly seen in the spectrum of V532 \citep{me}.  
\citet{me} estimated the size and the mass of the V532 nebula and concluded that parameters of the nebula are, by an order of magnitude, consistent with   
these typical for ejecta of LBV stars.  The emitting gas   
was probably ejected during one or several outburst events at wind   
velocities of about 100~$\kms$. The dust circumstellar envelope has not been detected \citep{Humphreys2014}.

\sectionb{5}{STUDIES OF SPECTRAL VARIABILITY AND SPECTRAL CLASSIFICATION}

      The first investigations of spectral variability were started by \citet{viotti2006,viotti2007}.  
      Using five spectra acquired in 2003 $\div$ 2006 \citet{viotti2006,viotti2007}  found an anti-correlation 
      between equivalent widths of the Wolf-Rayet blue bump at $4630\div 4686$~\AA \AA \ and the visual luminosity. 
      \citet{me} classified archived spectra of V532 and demonstrated an  evidence for a  correlation between 
      spectral type and visible magnitude. Also it was displayed that in the deep minimum (2008 year) the 
      spectrum of Romano's star  became similar to one of WN8 star type \citep{me}.

      It is easy enough to classify every single spectrum obtained at various
      moments of time, 
      but there is no consensus among the researchers on the type of the star. 
      As stated above, G.~Romano classified the star as a Hubble-Sandage variable. \citet{HumphreysDavidson} 
      on the basis of its variability classified the star as an LBV candidate. \citet{polcaro}  
      estimated the bolometric absolute magnitude of the object as $M_{bol} \approx -10\mag{.}4$, using bolometric
      correction ``of at least -3~mag'' and distance modulus $m-M=24\mag{.}8$. 
      They classify V532 as an LBV because the object fulfills all the criteria of \citet{HumphreysDavidson}. 

      An additional argument  for LBV/post-LBV status of V532 is its location in the galaxy.  
      V532 does not belong to any OB-association of M33 galaxy. 
      \citet{NathanSmith2014} concluded that  LBVs systematically avoid clusters of O-type
      stars, and they are almost never closely associated with O-type stars of similar (presumed) initial mass.


       However, \citet{polcaro10} concluded that bolometric luminosity has significantly 
       changed during minimum phase of 2008 year and suggested that Romano's star now is about to 
       end its LBV phase and to become a late WN type star. 

       In a recent article \citet{Humphreys2014} picked out one more feature of LBV star -- its slow stellar wind.
       LBVs have low wind speeds in their hot, quiescent or visual minimum state, compared to the B-type 
       supergiants and Of/WN stars which they spectroscopically resemble \citep{Humphreys2014}. 
       \citet{Humphreys2014} concluded that in spite of Romano's star having slow wind it is not an LBV star on all the characteristics. 
       Most probably  Romano's star has already passed its
       LBV phase, and is a post-LBV star now. 


\sectionb{6}{NUMERICAL SIMULATION}

              
\begin{table*}\centering
\caption{Derived properties of V532. H/He denotes hydrogen number fraction
  relative to helium, $f$ is a filling factor. 
} 
\label{tab:parmodel}
\bigskip
\begin{tabular}{lcccccccc}
\hline
Star                &      Sp.    &$T_{eff}$&$R_{2/3}$        & $\log L_*$      &   $ \dot{M}_{cl},10^{-5}$ & f  & $v_{\infty}$& H/He  \\
                    &      type   &   [kK]  &[$\rm R_{\odot}$]&[$\rm L_{\odot}$]&[$ \rm M_{\odot}\,yr^{-1}$]&    &[$\rm km/s$]&       \\
\hline
 2005$^1$           &       WN11h &  20.4   &  69.1           &  5.89           &    4.5                   & 0.5&  200        &  1.4      \\
 2008$^1$           &       WN8h  &  31.7   &  23.9           &   5.72          &    2                     & 0.1&  360    &1.9 \\
 2010$^2$           &      WN10h  &  26     &  41.5           &   5.85          &   2.18                   &0.25&  265        &1.5       \\ 
 2014               &       WN8h  &  32.7   &  22             &   5.72          &    1.7                   & 0.1&  400        &1.9 \\
\hline
\multicolumn{9}{l}{1 -- data taken from \citet{me2012pasha} and 2  -- from \citet{ClarkLBV2012}.}
\end{tabular}
\end{table*}

              Today numerical modeling provides most complete information about parameters of 
              stellar atmospheres. The first modeling of V532 atmosphere was performed by   \citet{me2012pasha}.
              Non-LTE radiative transfer code {\sc cmfgen} \citep{Hillier5} was used for the analysis.
              To accurately determine of the luminosity the model flux was recalculated for the distance of M33
              for every model. Distance to M33 was adopted as $D=847\pm60 \ \kpc$, which gives a distance modulus 
              of  $(m-M)=24.64\pm 0.15\magdot{\,}$ \citep{distance}. 
              Then, the simulated spectra were convolved with B- and V-band
              sensitivity curves. The resulting fluxes were converted 
              to magnitudes \citep{leng} and compared to the photometrical data. 
 
           \citet{me2012pasha} investigated the optical spectra of Romano's star in two different states, 
           the brightness minimum of 2008 (${\it B}=18.5 \pm 0.05$~mag) and a moderate
           brightening in 2005 (${\it B}=17.1\pm 0.03$~mag). 
           Figure~1 shows the observed spectra of V532 at different phases and the best-fit model spectra. 
           Stellar parameters derived for both models are given in Table~1.

       Main result of the work of \citet{me2012pasha} is that the  bolometric luminosities of V532 were
       different in 2005 and 2008. The bolometric luminosity of V532 in 2005 
       ($L_*=7.5\cdot10^{5}L_{\odot}$) is  1.5 times higher. This result confirms the conclusion of  \citet{polcaro10}.

       \citet{ClarkLBV2012} modelled the spectrum of Romano's star obtained in September 2010 when the
       V-band magnitude of the object was between 17$\mag{.}$75 and 17$\mag{.}$85. They also used {\sc cmfgen} code \citep{Hillier5}. 
       For comparison the results of \citet{ClarkLBV2012}  are also listed in Table~1. 
       \citet{ClarkLBV2012} inferred (see Table~1) 
       that the physical properties 
       ($T_{eff}$, $v_{\infty}$, filling factor $f$)
       of Romano's star in 2010 are a smooth progression between photometric minima and maxima. We assume that values of H/He agree  within the limits of errors. 
       It may be seen from the table that values of luminosities in 2008 and 2010 are different. 
       However, it must be noted that \citet{ClarkLBV2012} assumed that the distance to M33 is equal to 964~\kpc, 
       while \citet{me2012pasha} took  the distance of 847~\kpc. 
       If we recalculate the luminosity found by Clark et al.  $L_*=7\cdot10^{5}\Lsun$ 
       to distance 847~\kpc \ we get that luminosity of V532  is $L_*=5.4\cdot10^{5}\Lsun$ and it did not change after 2008 within the error limits ($L_*=5.2\pm0.2\cdot10^{5}\Lsun$,\ \citet{me2012pasha}). 
       
       
 
\begin{figure}[!tH]
\label{fig:HRdiagram1}
\vbox{
\centerline{\psfig{figure=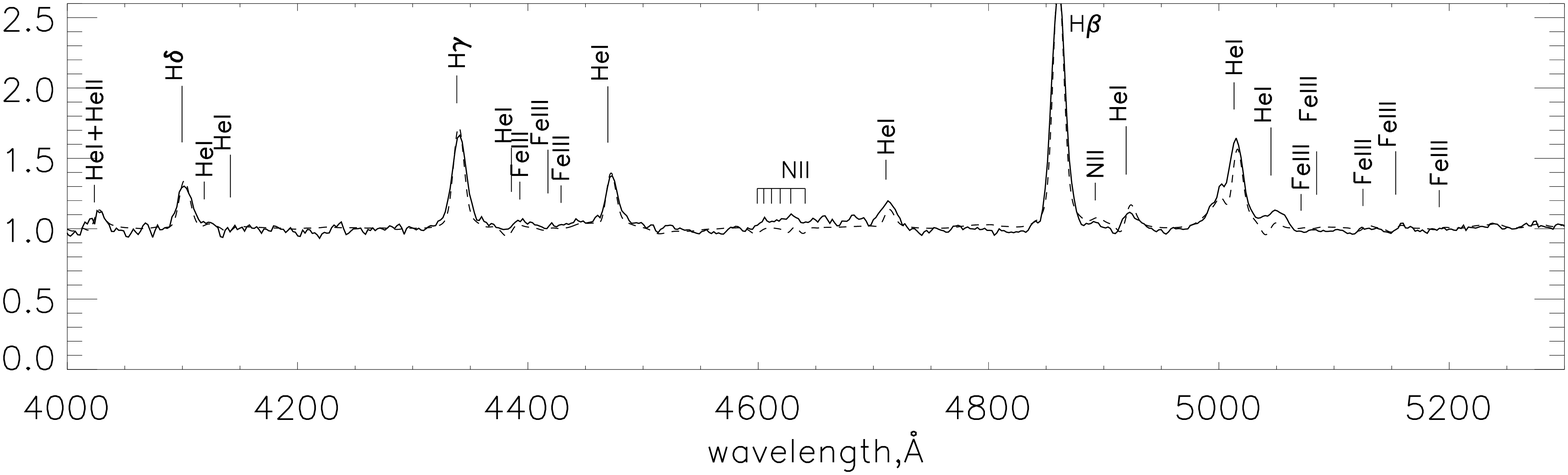,width=120mm,angle=0,clip=}}
\centerline{\psfig{figure=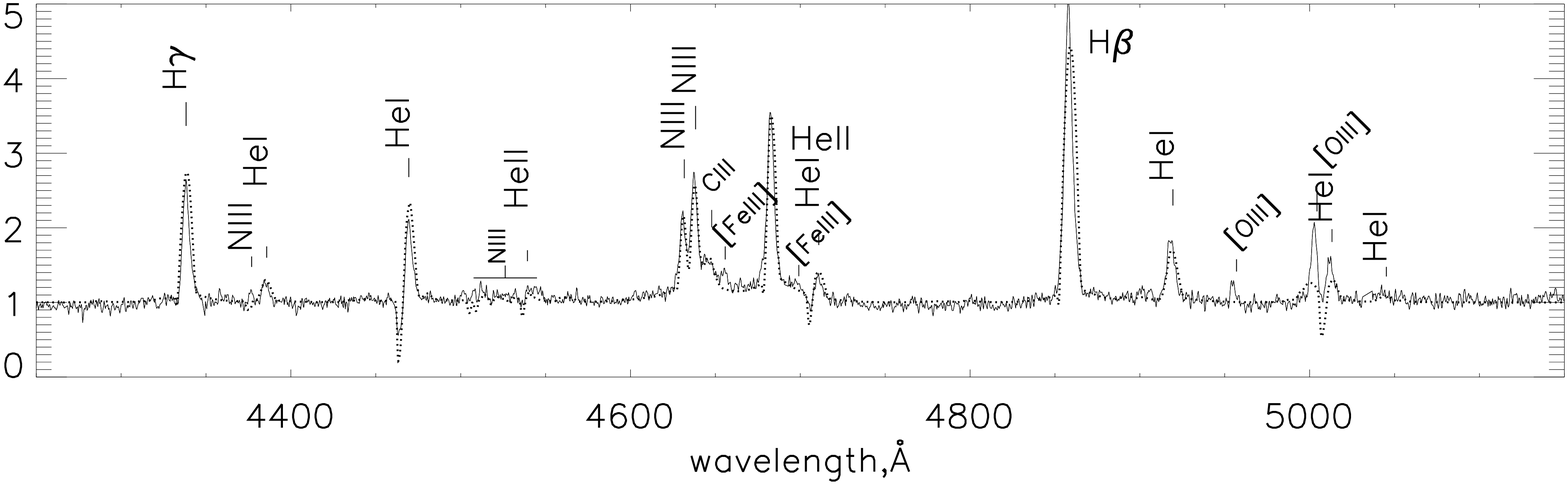,width=120mm,angle=0,clip=}}
\centerline{\psfig{figure=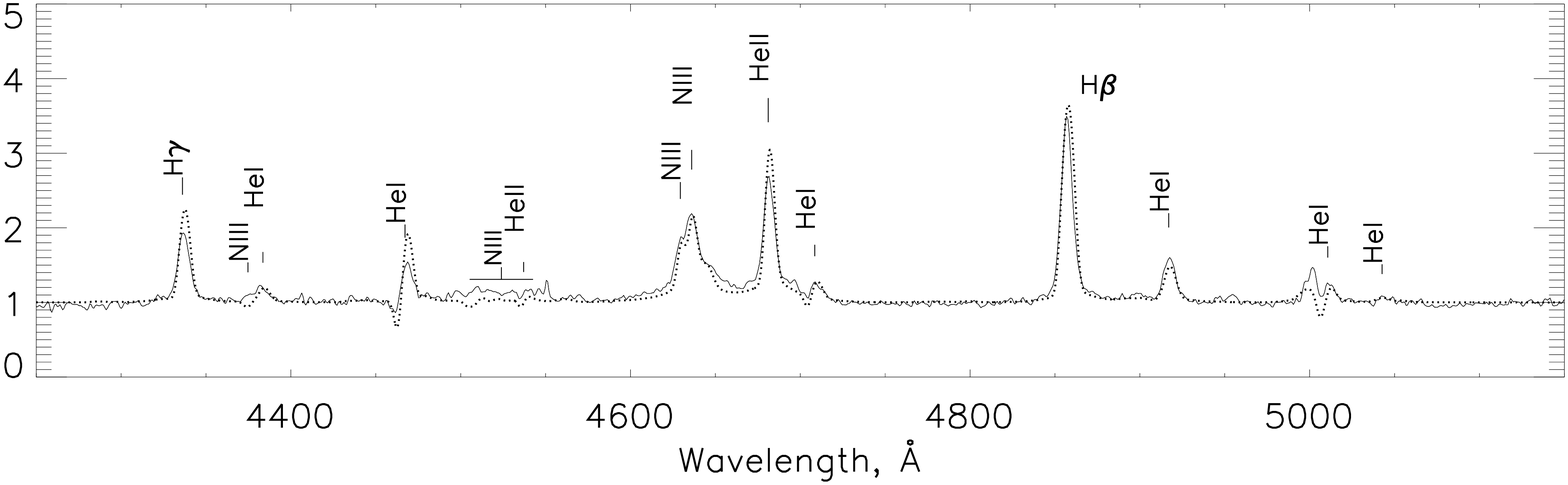,width=120mm,angle=0,clip=}}
\vspace{1mm}
\captionb{1}
{     The normalized optical spectra (solid line) 
      compared with the best-fit {\sc cmfgen} models (dashed line). 
      Top panel: the spectrum obtained  
      in February 2006 when V532 was $17\mag{.}27$ in V band. 
      Middle panel: the spectrum obtained in October 2007 (${\it V}=18\mag{\ .}68$). 
      Bottom panel: the spectrum obtained in August 2014 (${\it V}=18\mag{\ .}7$). 
      }
}
\end{figure}



 \sectionb{7}{NEW OBSERVATIONS}
 
        In August 2014 we have obtained a new spectrum of V532  on the Russian 6-m telescope 
        with the SCORPIO multi-mode focal reducer 
        in the long-slit mode~\citep{scorpio}. The spectrum was reduced using the {\tt ScoRe} 
        package written in {\sc idl} language. The package includes all the standard stages of 
        long-slit data reduction process. 
        
        Now V532 is in the minimum of brightness, its visible magnitude is V=18$\mag{.}$7~\footnote{ Photometric 
        data were kindly provided by Roberto Haver and Massimo Calabresi, scientists of Associazione Romana Astrofili (ARA)}
        Bottom panel of Figure~1 shows the spectrum obtained in August 2014 and its best-fit model. 
        As one may see from the figure, the new August spectrum is very similar to spectrum obtained in early 2008.  
        This is verification of hypothesis that the change of the stellar
        magnitude occurs simultaneously with spectral class changes. 
        The dependence of stellar magnitude on spectral class doesn't vary with time. Therefore we can use this property 
        for studies of the periods for which we have only photometric data.

 \sectionb{8}{CONCLUDING REMARKS}


          Figure~2 presents the positions of V532  in different phases in 
          the Hertzsprung--Russell  (HR) diagram according to modeling by \citet{me2012pasha}. 
          V532 in maximum of brightness ({\it V}=$17\mag{\,}$, Feb.~2005) 
          lies on the  LBV minimum instability strip. It moves to the ``forbidden region'' in the minimum of brightness. 
          V532 is the first star which demonstrated the transition from the instability strip to WR region \citep{meJenam}. 
          The main argument for what V532 is now not an LBV is that it does not exhibit S~Dor-like transitions to the cool, 
          dense wind state. Instead, it oscillates between two hot states on the HR  diagram  
          characterized by WN spectroscopic features \citep{Humphreys2014}. 
          However there is a spectrum obtained in 1992 which is classified as a late B spectral type \citep{szeifert}.
          Maybe this spectrum  is a confirmation that V532 in early 90s was actually in a cool state?

          Other interesting question is how did the luminosity of Romano's star change before and  after moderate
          brightening in 2005? How did other parameters of the atmosphere behave? 
          To answer these questions,  numerical modeling of archive spectra is needed.
          As mass loss rate  depends on luminosity it is necessary 
          to make   calculations for different phases in a consistent way.
          

          It is clearly seen from this review that V532 has been primarily studied in the optical range. 
          While its optical monitring is being performed regularly, in other wavebands it has not been studied well.
          Estimations of X-ray, ultraviolet and radio luminosities of V532 are necessary for precise   
          construction of its spectral energy distribution (SED).  Moreover it is interesting to observe 
          variability in different ranges. We consider that investigation of V532 in wide 
          range of wavelenghts is a top priority task. 
          

          
 \bigskip
 
  \thanks{I would like to thank Roberto Viotti, Roberto Haver and Massimo Calabresi for providing new photometric data and the Referee V.P. Arkhipova for valuable remarks given in the Referee report.  
  The study was supported by the Russian Foundation for Basic Research 
  (projects no. 12-07-00739-a, 14-02-31247, 14-02-00291). 
  I thank the Dynasty Foundation for a grant.}


\begin{figure}[!tH]
\label{fig:HRdiagram}
\vbox{
\centerline{\psfig{figure=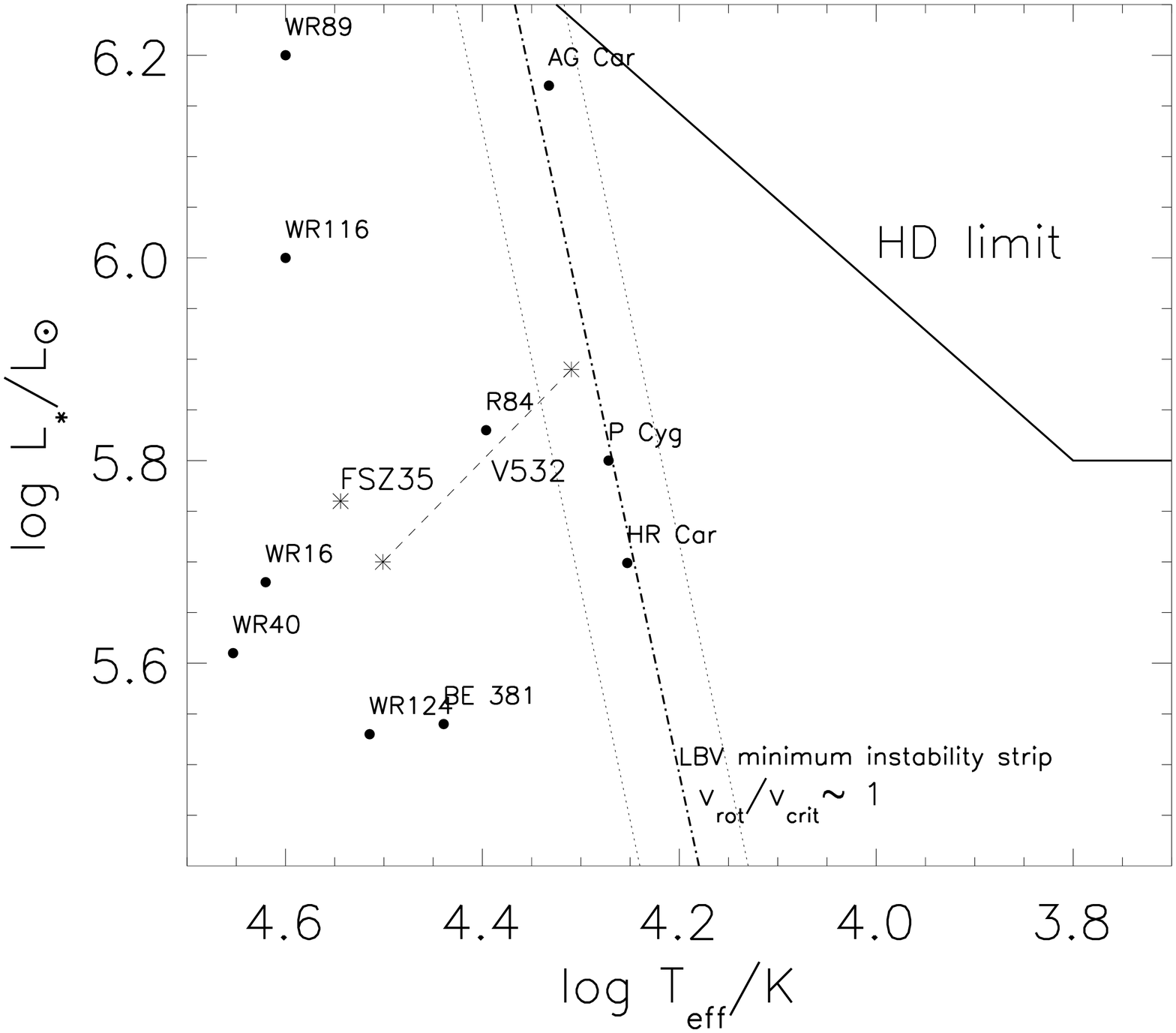,width=100mm,angle=0,clip=}}
\vspace{1mm}
\captionb{2}
{       
         A schematic of upper part of HR Diagram. 
         Position for the LBV minimum instability strip is provided by dashed-dotted line.
         The location of the Humphreys-Davidson limit \citep{HumphreysDavidson} 
         is shown by solid line. Dashed line shows transit V532 across instability strip.
         WN9 (BE381, R84), WN8 (WR124, WR16, WR40, FSZ35) 
         and LBV (AG~Car, HR~Car, P~Cyg) stars are shown for comparison. 
         Data on these objects were taken from  \citep{CrowtherLMC}, \citep{CrowtherLMC}, 
         \citep{Crowther99}, \citep{HHH}, \citep{HHH}, \citep{fsz35}, \citep{groh}, 
         \citep{grohHR}, \citep{NajarroPCyg}, consequently. }
}
\end{figure}


        \end{document}